\documentclass[12pt,a4paper]{article}
\usepackage{graphicx}
\usepackage{amsmath,amssymb,amsthm}
\usepackage{hyperref}
\usepackage{xcolor}
\usepackage{cite}
\usepackage[left=2cm,right=2cm,top=2.5cm,bottom=2.5cm]{geometry}

\newcommand{\lsim}{\lesssim}

\newcommand{\delEW}{\Delta_{\mathrm{EW}}}

\def\eslt{\not\!\!{E_T}}
\def\to{\rightarrow}

\def\bi{\begin{itemize}}
\def\ei{\end{itemize}}

\def\tchi{\tilde\chi}

\def\sps1ap{SPS1a$^\prime$}
\def\c1p{C1$^\prime$}

\def\tg{\tilde g}

\def\agt{\gtrsim}
\def\be{\begin{equation}}  
\def\ee{\end{equation}}  
\def\bea{\begin{eqnarray}}  
\def\eea{\end{eqnarray}}  
\def\beas{\begin{eqnarray*}}  

\begin{document}
\begin{center}
  {\Large \bf Supersymmetry at High Luminosity LHC}\\
\vspace{0.3cm} \renewcommand{\thefootnote}{\fnsymbol{footnote}}
{\large Kairui Zhang$^{1}$\footnote[1]{Email: kzhang25@ou.edu}\footnote[2]{Presented at the 32nd International Symposium on Lepton Photon Interactions at High Energies,
Madison, Wisconsin, USA, August 25--29, 2025}
}\\ 
\vspace{0.2cm} \renewcommand{\thefootnote}{\arabic{footnote}}
{\it 
$^1$Department of Physics and Astronomy,
University of Oklahoma, Norman, OK 73019, USA \\
}
\date{}
\end{center}

\begin{abstract}
\noindent
Weak-scale supersymmetry (SUSY) is well motivated as a technically natural solution to the gauge hierarchy problem. LHC limits on superpartners, however, have sharpened the Little Hierarchy problem, raising the question of why $m_{\rm weak}\ll m_{\rm soft}$. We review current collider and dark matter constraints and their implications for leading SUSY-breaking scenarios. Electroweak naturalness is revisited using a weak-scale measure that avoids ambiguities associated with high-scale fine-tuning. Also, within the string landscape, soft terms are statistically favored to be large, while anthropic selection enforces a weak scale near the observed value. This framework—often termed \emph{stringy naturalness}—naturally accommodates $m_h\simeq125$~GeV while placing sparticles above current LHC limits. Updated HL-LHC projections for non-universal Higgs mass (NUHM) models—the most plausible realization consistent with above picture—show that searches for higgsinos, stops, and heavy Higgs bosons will soon begin to probe the core of the viable parameter space.
\end{abstract}


\section{Introduction}
\label{sec:intro}
The Standard Model (SM) successfully describes all established collider data. The discovery of a Higgs boson with mass $m_h\simeq125$~GeV confirmed the mechanism of electroweak symmetry breaking (EWSB) via a Higgs vacuum expectation value (VEV) and completed the SM spectrum. Yet, the SM is not fundamental. It suffers from three fine-tuning problems: the cosmological constant problem, the strong CP problem, and the gauge hierarchy problem (GHP). The most plausible solutions are, respectively, anthropic selection of a small vacuum energy in a multiverse, the Peccei–Quinn mechanism with its associated axion, and weak-scale supersymmetry (WSS). Under $R$-parity conservation, SUSY predicts a stable lightest supersymmetric particle (LSP), a weakly interacting massive particle (WIMP), as a dark matter candidate.

WSS is also supported indirectly by several loop-level effects. For soft terms $m_{\rm soft}\sim{\rm TeV}$, the large top Yukawa coupling naturally triggers radiative EWSB~\cite{Ibanez:1982fr,Alvarez-Gaume:1983drc}. The MSSM improves gauge coupling unification~\cite{Dimopoulos:1981yj} and predicts an upper bound $m_h\lesssim130$~GeV for the light Higgs boson~\cite{Carena:2002es}. Electroweak precision data also prefer heavy SUSY over the SM~\cite{Heinemeyer:2006px}. Despite these motivations, direct LHC searches have not yet revealed superpartners. Most analyses employ simplified models~\cite{LHCNewPhysicsWorkingGroup:2011mji}, which overstate the true reach since realistic SUSY spectra involve multiple production channels and cascade decays~\cite{Baer:1986au,Drees:2013wra,Kraml:2013mwa,GAMBIT:2017yxo}. 

In most MSSM studies, $R$-parity conservation is imposed by hand to forbid rapid proton decay. Discrete gauged $R$-symmetries $\mathbb{Z}_{n}^R$ arising from string compactification can instead forbid the dangerous renormalizable RPV operators and simultaneously generate the Peccei–Quinn symmetry~\cite{Baer:2018avn}. Recent work~\cite{Baer:2025oid,Baer:2025srs} shows that implementing a $\mathbb{Z}_{4,8}^R$, non-renormalizable operators induce small effective RPV couplings, which preserve collider $\eslt$ signatures but allow relic WIMPs to decay before BBN, leaving an axion-dominated cold dark matter universe.

Ton-scale noble-liquid detectors such as LZ~\cite{LZ:2024zvo} have now excluded well-tempered neutralinos and most bino-like WIMPs by several orders of magnitude. Spin-dependent and indirect searches further disfavor wino-like dark matter. Thermally produced higgsino-like WIMPs would also be excluded if they comprised all of the dark matter. In models with mixed axion/WIMP dark matter, however, non-thermal processes or small $R$-parity-violating decays can alter the relic abundance, keeping such scenarios consistent with present bounds.

\section{Electroweak naturalness and landscape expectations}
\label{sec:nat}
\subsection{Electroweak naturalness measure}
The absence of superpartners at the LHC has sometimes been taken as a sign that SUSY is unnatural. Earlier estimates based on high-scale sensitivity measures tended to exaggerate fine-tuning because they neglected parameter correlations~\cite{Ellis:1986yg,Barbieri:1987fn}. A more conservative and model-independent criterion is obtained from the weak-scale Higgs potential minimization condition:
\begin{equation}
    \frac{m_Z^2}{2} = \frac{m_{H_d}^2+\Sigma_d^d-(m_{H_u}^2+\Sigma_u^u)\tan^2\beta}{\tan^2\beta -1} - \mu^2 ,
\label{eq:mzs}
\end{equation}
where $\Sigma_{u,d}^{u,d}$ denote the one- and two-loop corrections. We define
\begin{equation}
    \Delta_{EW}\equiv \frac{\max|\text{terms on RHS of Eq.~\ref{eq:mzs}}|}{m_Z^2/2} ,
\label{eq:dew}
\end{equation}
and regard a model as natural if $\Delta_{EW}\lesssim 30$~\cite{Baer:2012up,Baer:2012cf,Mustafayev:2014lqa}.  
This directly connects fine-tuning to weak-scale parameters without reference to high-scale inputs. The upper bounds inferred from $\Delta_{EW}$ are (i)$\mu\lesssim350$~GeV, leading to light higgsinos; (ii) stops are at most a few TeV range and highly mixed since $\Sigma_u^u(\tilde t_{1,2})$ is then reduced; (iii) gluinos up to $\sim6$~TeV and first/second generation squarks/sleptons $\lsim40$~TeV are allowed. Hence present LHC limits are compatible with electroweak natural SUSY.

\subsection{Landscape expectation}
In the string landscape, the values of SUSY-breaking $F$- and $D$-term VEV can vary across different vacua. The resulting distribution of soft terms then follows a power law,
\begin{equation}
    f_{\rm SUSY}\sim m_{\rm soft}^n,\qquad n=2n_F+n_D-1 ,
    \label{eq:land}
\end{equation}
so that vacua with larger soft terms are statistically favored~\cite{Douglas:2004zg,Douglas:2004qg}. Yet, anthropic selection then requires the derived weak scale in each pocket universe to lie within the ABDS window~\cite{Agrawal:1997gf},
\begin{equation}
    0.5\,m_{weak}^{OU} < m_{weak}^{PU} < (2\text{--}5)\,m_{weak}^{OU}.
\end{equation}
Vacua predicting much larger weak scales are excluded, so the remaining ones cluster near the largest soft terms still compatible with weak scale close to the observed scale. When mapped to low-energy MSSM spectrum: (i) first/second-generation scalars with masses of tens of~TeV, naturally solving SUSY flavor and CP problem via a quasi-degenerate and decoupling way, (ii) third-generation scalar at the few-TeV scale with large mixing -- exactly what lifting the light Higgs$\sim 125$ GeV, (iii) gluino at TeV scale, (iv) a small $\mu$ term near the weak scale.

\section{Update on dark matter in natural SUSY models}
\label{sec:DM}
Natural SUSY models with a thermally underproduced higgsino LSP now appear excluded by the LZ2024 limits on spin-independent WIMP–nucleon scattering~\cite{LZ:2024zvo}. However, SUSY—especially string-motivated—frameworks contain several non-thermal mechanisms that can alter the neutralino relic abundance or even replace WIMPs with other dark matter components~\cite{Baer:2025oid,Baer:2025srs}.

In natural SUSY, the axionic solution to the strong CP problem arises naturally. The supersymmetrized DFSZ model includes a superpotential term
\begin{equation}
    W \ni \lambda_\mu \frac{S^2 H_u H_d}{m_P},
\end{equation}
which solves the $\mu$ problem via the Kim–Nilles mechanism~\cite{Kim:1983dt} when the PQ charged field $S$ develops a VEV of order the PQ scale
$f_a\sim 10^{11}$ GeV: then a weak scale $\mu$ term develops
with 
\begin{equation}
    \mu = \lambda_\mu \frac{f_a^2}{m_P} \sim m_{weak}.
\end{equation}
SUSY also stabilizes the PQ symmetry against higher-order operators, addressing the axion quality problem. Discrete gauged $R$-symmetries such as $\mathbb{Z}_{4,6,8,12,24}^R$~\cite{Nilles:2017heg,Lee:2011dya,Baer:2018avn,Bhattiprolu:2021rrj} can forbid renormalizable RPV, proton-decay operators, and $\mu$ term at Planck scale. Under $\mathbb{Z}_{24}^R$, PQ violation arises only through operators of dimension $\geq 11$, ensuring a high-quality axion. The same discrete $R$-symmetry involving PQ-charged hidden fields can regenerate tiny effective RPV couplings~\cite{Baer:2025oid,Baer:2025srs} and weak-scale $\mu$ via Kim-Nilles above. These can cause relic higgsinos to decay before BBN, resulting in an axion-only cold dark matter universe.

\section{SUSY models on the plausibility meter}
\label{sec:models}
Given current LHC bounds, $m_h\simeq125$~GeV, and the requirement of electroweak naturalness $\Delta_{EW}\lesssim30$, SUSY constructions can be divided into \textit{implausible} models—those requiring hidden fine-tuning—and \textit{plausible} ones that naturally compatible with collider limits.

\begin{table}
\centering
\resizebox{0.8\columnwidth}{!}{%
\begin{tabular}{lcccccc}
\hline
model & $\tilde{m}(1,2)$ & $\tilde{m}(3)$ & gauginos & higgsinos & $m_h$ & $P_\mu$ \\
\hline
SM & - & - & - & -& - & $7\cdot10^{-27}$ \\
CMSSM ($\Delta_{EW}=2641$) & $\sim 1$ & $\sim 1$ & $\sim 1$ & $\sim 1$ & $0.1-0.13$
& $5\cdot 10^{-3}$ \\
PeV SUSY & $\sim 10^3$ & $\sim 10^3$ & $\sim 1$ & $1-10^3$ &
$0.125-0.155$ & $5\cdot 10^{-6}$ \\
Split SUSY & $\sim 10^6$ & $\sim 10^6$ & $\sim 1$ & $\sim 1$ & $0.13-0.155$
& $7\cdot 10^{-12}$ \\
HS-SUSY & $\agt 10^2$ & $\agt 10^2$ & $\agt 10^2$ & $\agt 10^2$ & $0.125-0.16$
& $6\cdot 10^{-4}$ \\
Spread ($\tilde{h}$LSP) & $10^{5}$  & $10^5$ & $10^2$ & $\sim 1$ & $0.125-0.15$ & $9\cdot 10^{-10}$ \\
Spread ($\tilde{w}$LSP) & $10^{3}$ & $10^{3}$ & $\sim 1$ & $\sim 10^2$ & $0.125-0.14$  & $5\cdot 10^{-6}$ \\
Mini-Split ($\tilde{h}$LSP)& $\sim 10^4$ & $\sim 10^4$ & $\sim 10^2$ & $\sim 1$  & $0.125-0.14$ & $8\cdot10^{-8}$ \\
Mini-Split ($\tilde{w}$LSP)& $\sim 10^2$ & $\sim 10^2$ & $\sim 1$ & $\sim 10^2$ & $0.11-0.13$ & $4\cdot 10^{-4}$ \\
SUN-SUSY  & $\sim 10^2$ & $\sim 10^2$ & $\sim 1$ & $\sim 10^2$  & $0.125$
& $4\cdot 10^{-4}$ \\
G$_2$MSSM  & $30-100$ & $30-100$ & $\sim 1$  & $\sim 1$  & $0.11-0.13$
& $2\cdot 10^{-3}$ \\
RNS/landscape & $5-40$  & $0.5-3$ & $\sim 1$ & $0.1-0.35$ & $0.123-0.126$
& $1.4$ \\
\hline
\end{tabular}}
\caption{A survey of some unnatural and natural SUSY models
  along with general expectations for sparticle and Higgs
  mass spectra in TeV units. Relative probability measure $P_\mu$ for the model to emerge from the landscape is also shown. The table is taken from Ref.~\cite{Baer:2025zqt}.
}
\label{tab:models}
\end{table}

\subsection{Implausible models.}
CMSSM/mSUGRA, GMSB, minimal AMSB relying on heavy stops, due to tiny trilinear $A$-terms, to reproduce $m_h\simeq125$~GeV, giving $\Delta_{EW}\gg30$~\cite{Baer:2014ica,Baer:2024fgd}. Likewise, split, PeV, or high-scale SUSY spectra abandon naturalness and are statistically disfavored in the landscape. Gaugino-mediated (inoMSB)/no-scale SUGRA has similar situation~\cite{Baer:2024fgd} and often featured by charged LSP.

\subsection{Plausible: NUHM2–4, natural AMSB, generalized mirage mediation~\cite{Baer:2020dri,Baer:2024hpl,Baer:2018hwa,Baer:2016hfa,Baer:2024tfo}}

\textbf{Gravity-mediation described by NUHM models:}
once we allow $m_{H_u}^2$ and $m_{H_d}^2$ to differ from the matter scalars—as expected from generic SUGRA/string compactifications—one can choose $\mu \sim 100$–350~GeV, obtain $m_h\simeq125$~GeV from large $A_0$, and keep $m_{\tg}\sim 2.5$–6~TeV and third-generation squarks at a few TeV, all with $\Delta_{EW}\lesssim 30$. The landscape pull to large but not too large soft terms then favors precisely this corner, with first/second generation scalars in the multi–tens–of–TeV range.

\textbf{Natural AMSB (nAMSB):}
if AMSB is supplemented by non-universal bulk scalar masses and bulk $A$-terms (as is reasonable in extra-dimensional or string setups), one can lower $\mu$, raise $m_h$ with large $A$-terms, and keep the wino heavier than the higgsino, yielding an EW-natural spectrum still consistent with current LHC wino and gluino bounds.

\textbf{Natural generalized mirage mediation (nGMM/GMM):}
when the mirage parameters are treated as continuous—as is appropriate for generic Calabi–Yau compactifications—the same pattern emerges: small $\mu$, large (but tuned-by-landscape) soft terms, $m_h\simeq125$~GeV from sizable $A$-terms, and gaugino masses unifying at a mirage scale. These models sit comfortably in the LHC-allowed, low-$\Delta_{EW}$ region.

\subsection{Lessons}
Imposing (i) $m_h\simeq 125$~GeV, (ii) present LHC limits on sparticles, and (iii) $\Delta_{EW}\lesssim 30$ effectively filters out the historically popular minimal models and leaves a small, well-motivated set—NUHM-type gravity mediation, nAMSB, and nGMM—in which light higgsinos are guaranteed while the rest of the spectrum can be heavy. This is also the pattern preferred by landscape arguments that cap the pocket-universe weak scale near the observed value. A survey across different SUSY model is summarized in Table~\ref{tab:models}.

\section{HL-LHC phenomenology of natural SUSY}
\label{sec:hllhc}
We collect here the HL-LHC ($pp$ at $\sqrt{s}=14$~TeV, 3~ab$^{-1}$) channels that are sensitive to natural, landscape-motivated spectra.

\paragraph{Higgsino pair production.}
ISR-assisted dilepton + jet + $\eslt$ searches can probe $m_{\tilde\chi_2^0}$ up to $\sim 300$~GeV for moderate mass gaps, covering most natural higgsino scenarios~\cite{Han:2014kaa,Baer:2014kya,Baer:2021srt}.

\paragraph{Top-squark pair production.}
For $\tilde t_1\to t\tilde\chi_{1,2}^0$ and $b\tilde\chi_1^+$, HL-LHC reaches up to $\sim 1.7$~TeV for discovery and $\sim 2$~TeV for exclusion~\cite{Baer:2023uwo}, covering much (but not all) of the parameter space favored by landscape-selected natural SUSY.

\paragraph{Wino pair production.}
Combined reach of various channels can extend to $M_2\simeq 1.1$~TeV (5$\sigma$) and $\sim1.4$~TeV (95\%~CL)~\cite{Baer:2023olq}, probing roughly half of the wino mass range favored by the landscape.

\paragraph{Gluino pair production.}
For $\tilde g\to t\tilde t_1$ leading to multi-$b$ + $\eslt$, followed by $\tilde t_1\to b\tchi_1^+$ or $\tilde t_1\to t\tchi_{1,2}^0$, HL-LHC can reach $m_{\tilde g}\sim 2.8$~TeV~\cite{Baer:2016wkz}, testing the lower end of the 2–6~TeV range expected from landscape considerations.

\paragraph{Stau pair production.}
The process $pp\to\tilde\tau_R^+\tilde\tau_R^-\to\tau^+\tau^-+\eslt$ has been studied in both $\tau_h\tau_h$ and $\tau_h\ell$ final states, allowing 95\%~CL exclusions up to several hundred GeV, though no discovery reach is expected~\cite{Baer:2024hgq}.

\begin{figure}[htb!]
\begin{center}
  \includegraphics[height=0.35\textheight]{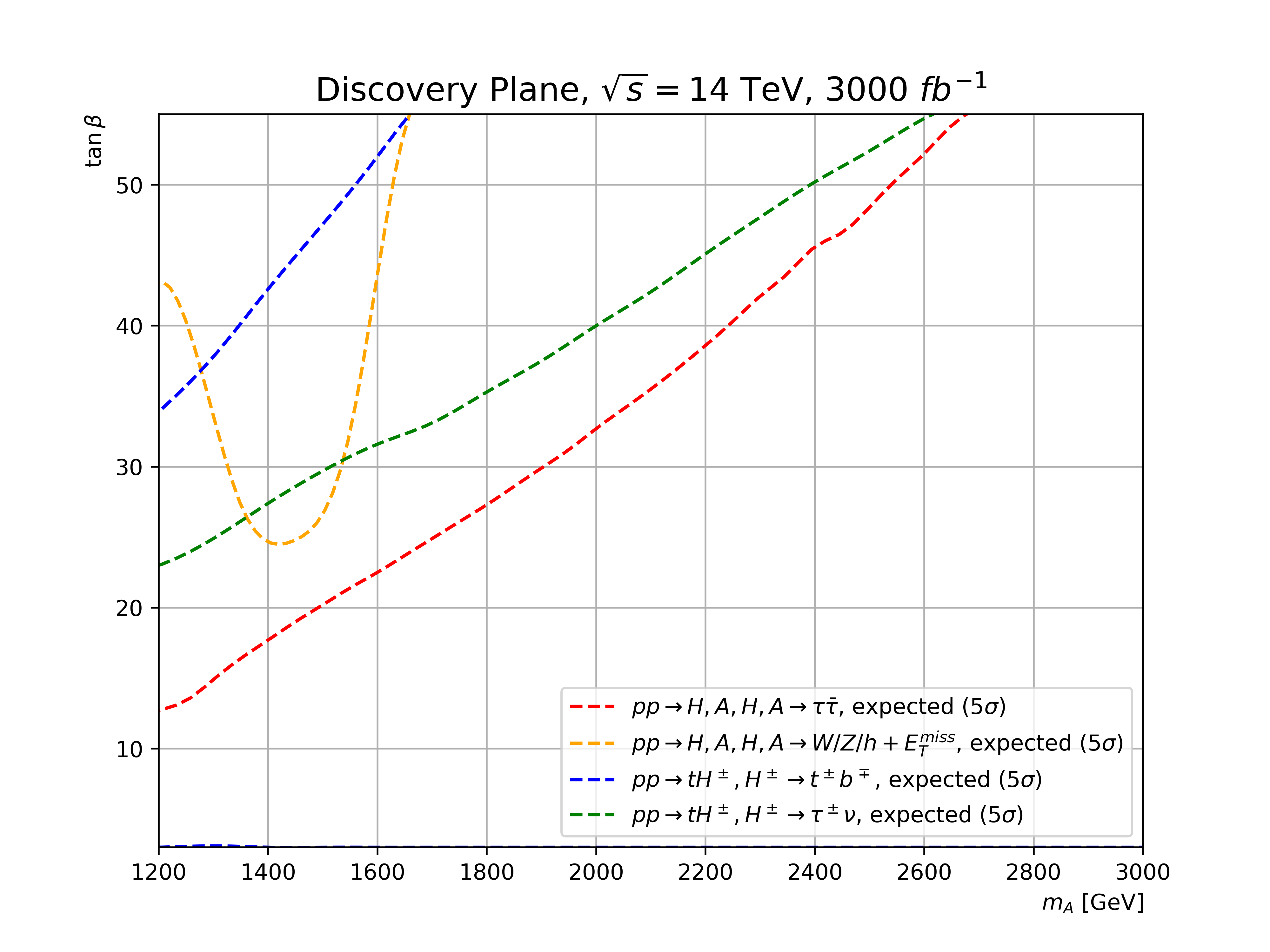}
  \caption{Plot of $5\sigma$ discovery projections for
    heavy SUSY Higgs boson searches at HL-LHC
    in the $m_A$ vs. $\tan\beta$ plane for the $m_h^{125}({\rm nat})$ scenario.}
\label{fig:disc}
\end{center}
\end{figure}

Heavy Higgs searches in natural SUSY include both the standard $A,H\to\tau^+\tau^-$ mode and SUSY channels such as $A,H\to\tilde\chi\tilde\chi$~\cite{Baer:2022qqr,Baer:2022smj}. Charged Higgs bosons have been examined in $H^\pm\to tb$ and $\tau\nu$ decays~\cite{Baer:2023yxk}. The combined HL-LHC sensitivity is summarized in Fig.~\ref{fig:disc}.

\section{Conclusions}
SUSY remains the most coherent extension of the SM that stabilizes the weak scale and accommodates the observed Higgs mass. The electroweak measure $\delEW$ provides a conservative and model-independent way to assess naturalness, predicting light higgsinos, moderately heavy stops and gluinos, much heavier first/second generation squarks/sleptons. Except for the first/second generation squarks/sleptons, all natural SUSY partners and Heavy Higgs states can be probed in a future muon collider as well~\cite{Baer:2025uzx}. Landscape considerations reinforce this expectation statistically. The HL-LHC will test this paradigm in near future, and explore the simplest natural solutions to the hierarchy problem.

\vspace{0.5em}
\noindent\textbf{Acknowledgments.}
The author acknowledges support from the Homer L.Dodge Department of Physics and Astronomy at the University of Oklahoma and the Avenir Foundation. The author thanks his collaborators H.Baer, V.Barger, J.Bolich, J.Dutta, D.Martinez, S.Salam, and D.Sengupta for the work Ref.~\cite{Baer:2025zqt} this talk is based on.

\bibliography{ref}
\bibliographystyle{elsarticle-num}

\end{document}